\documentclass[journal]{IEEEtran}
\usepackage{flushend} 
\usepackage{lineno}
\usepackage{cite}
\usepackage{amsmath,amssymb}
\usepackage{tabu}
\usepackage{array}
\usepackage{comment}

\usepackage[nolist]{acronym}
\usepackage{acronym}

\ifCLASSOPTIONcompsoc
    \usepackage[caption=false, font=normalsize, labelfont=sf, textfont=sf]{subfig}
\else
\usepackage[caption=false, font=footnotesize]{subfig}
\fi

\ifCLASSINFOpdf
\else
\fi

 \usepackage{bibentry} 
\usepackage{graphicx}
\usepackage{epstopdf}
\usepackage{color}
\usepackage{amsmath}
\usepackage{amssymb}
\usepackage{amsfonts}
\usepackage{comment}
\usepackage{lipsum}
\usepackage{mathtools}
\usepackage{cuted}
\usepackage[pdftex]{hyperref}

\usepackage{algorithmic}

\usepackage{array}
\usepackage{mdwmath}
\usepackage{mdwtab}
\usepackage{eqparbox}

 \DeclarePairedDelimiterX\MeijerM[3]{\lparen}{\rparen}%
{\begin{smallmatrix}#1 \\ #2\end{smallmatrix}\delimsize\vert\,#3}

\newcommand\Gfun[8][]{%
  G^{\,#2,#3}_{#4,#5}\MeijerM[#1]{#6}{#7}{#8}}

\def\Poutasy{\textrm{P}_{out}^{asy}}

\WithSuffix\newcommand\Gfun*[7]{%
  G^{\,#1,#2}_{#3,#4}\MeijerM*{#5}{#6}{#7}}
\def\Pe{{P_e}}

\def\Peasy{{{P_e}^{asy}}}

\makeatletter
\DeclareRobustCommand\bfseriesitshape{%
\not@math@alphabet\itshapebfseries\relax
\fontseries\bfdefault
\fontshape\itdefault
\selectfont
}
\makeatother

\DeclareTextFontCommand{\textbfit}{\bfseriesitshape}

\begin{document}
\begin{acronym}

\acro{5G-NR}{5G New Radio}
\acro{3GPP}{3rd Generation Partnership Project}
\acro{AC}{address coding}
\acro{ACF}{autocorrelation function}
\acro{ACR}{autocorrelation receiver}
\acro{ADC}{analog-to-digital converter}
\acrodef{aic}[AIC]{Analog-to-Information Converter}     
\acro{AIC}[AIC]{Akaike information criterion}
\acro{aric}[ARIC]{asymmetric restricted isometry constant}
\acro{arip}[ARIP]{asymmetric restricted isometry property}

\acro{ARQ}{automatic repeat request}
\acro{AUB}{asymptotic union bound}
\acrodef{awgn}[AWGN]{Additive White Gaussian Noise}     
\acro{AWGN}{additive white Gaussian noise}

\acro{PU}{primary user}
\acro{SU}{secondary user}

\acro{APSK}[PSK]{asymmetric PSK} 
\acro{waric}[AWRICs]{asymmetric weak restricted isometry constants}
\acro{warip}[AWRIP]{asymmetric weak restricted isometry property}
\acro{BCH}{Bose, Chaudhuri, and Hocquenghem}        
\acro{BCHC}[BCHSC]{BCH based source coding}
\acro{BEP}{bit error probability}
\acro{BFC}{block fading channel}
\acro{BG}[BG]{Bernoulli-Gaussian}
\acro{BGG}{Bernoulli-Generalized Gaussian}
\acro{BPAM}{binary pulse amplitude modulation}
\acro{BPDN}{Basis Pursuit Denoising}
\acro{BPPM}{binary pulse position modulation}
\acro{BPSK}{binary phase shift keying}
\acro{BPZF}{bandpass zonal filter}
\acro{BU}[BU]{Bernoulli-uniform}
\acro{BER}{bit error rate}
\acro{BS}{base station}
\acro{BC}{backscatter communications}
\acro{SER}{symbol error rate}
\acro{CP}{Cyclic Prefix}
\acrodef{cdf}[CDF]{cumulative distribution function}   
\acro{CDF}{cumulative distribution function}
\acrodef{c.d.f.}[CDF]{cumulative distribution function}
\acro{CCDF}{complementary cumulative distribution function}
\acrodef{ccdf}[CCDF]{complementary CDF}               
\acrodef{c.c.d.f.}[CCDF]{complementary cumulative distribution function}
\acro{CD}{cooperative diversity}

\acro{CDMA}{Code Division Multiple Access}
\acro{ch.f.}{characteristic function}
\acro{CIR}{channel impulse response}
\acro{cosamp}[CoSaMP]{compressive sampling matching pursuit}
\acro{CR}{cognitive radio}
\acro{cs}[CS]{compressed sensing}                   
\acrodef{cscapital}[CS]{Compressed sensing} 
\acrodef{CS}[CS]{compressed sensing}
\acro{CSI}{channel state information}
\acro{CCSDS}{consultative committee for space data systems}
\acro{CC}{convolutional coding}
\acro{Covid19}[COVID-19]{Coronavirus disease}
\acro{SSBC}{spectrum sharing backscatter communications}
\acro{CW}{continuous wave}

\acro{DAA}{detect and avoid}
\acro{DAB}{digital audio broadcasting}
\acro{DCT}{discrete cosine transform}
\acro{dft}[DFT]{discrete Fourier transform}
\acro{DR}{distortion-rate}
\acro{DS}{direct sequence}
\acro{DS-SS}{direct-sequence spread-spectrum}
\acro{DTR}{differential transmitted-reference}
\acro{DVB-H}{digital video broadcasting\,--\,handheld}
\acro{DVB-T}{digital video broadcasting\,--\,terrestrial}
\acro{DL}{downlink}
\acro{DSSS}{Direct Sequence Spread Spectrum}
\acro{DFT-s-OFDM}{Discrete Fourier Transform-spread-Orthogonal Frequency Division Multiplexing}
\acro{DAS}{distributed antenna system}
\acro{DNA}{Deoxyribonucleic Acid}

\acro{EC}{European Commission}
\acro{EED}[EED]{exact eigenvalues distribution}
\acro{EIRP}{Equivalent Isotropically Radiated Power}
\acro{ELP}{equivalent low-pass}
\acro{eMBB}{enhanced mobile broadband}
\acro{EMF}{electric and magnetic fields}
\acro{EU}{European union}

\acro{FC}[FC]{fusion center}
\acro{FCC}{Federal Communications Commission}
\acro{FEC}{forward error correction}
\acro{FFT}{fast Fourier transform}
\acro{FH}{frequency-hopping}
\acro{FH-SS}{frequency-hopping spread-spectrum}
\acrodef{FS}{Frame synchronization}
\acro{FSsmall}[FS]{frame synchronization}  
\acro{FDMA}{Frequency Division Multiple Access}    

\acro{FD}{full-duplex}

\acro{GA}{Gaussian approximation}
\acro{GF}{Galois field }
\acro{GG}{Generalized-Gaussian}
\acro{GIC}[GIC]{generalized information criterion}
\acro{GLRT}{generalized likelihood ratio test}
\acro{GPS}{Global Positioning System}
\acro{GMSK}{Gaussian minimum shift keying}
\acro{GSMA}{Global System for Mobile communications Association}

\acro{HAP}{high altitude platform}

\acro{IDR}{information distortion-rate}
\acro{IFFT}{inverse fast Fourier transform}
\acro{iht}[IHT]{iterative hard thresholding}
\acro{i.i.d.}{independent, identically distributed}
\acro{IoT}{Internet of Things}                      
\acro{IR}{impulse radio}
\acro{lric}[LRIC]{lower restricted isometry constant}
\acro{lrict}[LRICt]{lower restricted isometry constant threshold}
\acro{ISI}{intersymbol interference}
\acro{ITU}{International Telecommunication Union}
\acro{ICNIRP}{International Commission on Non-Ionizing Radiation Protection}
\acro{IEEE}{Institute of Electrical and Electronics Engineers}
\acro{ICES}{IEEE international committee on electromagnetic safety}
\acro{IEC}{International Electrotechnical Commission}
\acro{IARC}{International Agency on Research on Cancer}
\acro{IS-95}{Interim Standard 95}

\acro{LEO}{low earth orbit}
\acro{LF}{likelihood function}
\acro{LLF}{log-likelihood function}
\acro{LLR}{log-likelihood ratio}
\acro{LLRT}{log-likelihood ratio test}
\acro{LOS}{Line-of-Sight}
\acro{LRT}{likelihood ratio test}
\acro{wlric}[LWRIC]{lower weak restricted isometry constant}
\acro{wlrict}[LWRICt]{LWRIC threshold}
\acro{LPWAN}{low power wide area networks}
\acro{LoRaWAN}{low power long range wide area network}
\acro{NLOS}{non-line-of-sight}

\acro{MB}{multiband}
\acro{MC}{multicarrier}
\acro{MDS}{mixed distributed source}
\acro{MF}{matched filter}
\acro{m.g.f.}{moment generating function}
\acro{MI}{mutual information}
\acro{MIMO}{multiple-input multiple-output}
\acro{MISO}{multiple-input single-output}
\acrodef{maxs}[MJSO]{maximum joint support cardinality}                       
\acro{ML}[ML]{maximum likelihood}
\acro{MMSE}{minimum mean-square error}
\acro{MMV}{multiple measurement vectors}
\acrodef{MOS}{model order selection}
\acro{M-PSK}[${M}$-PSK]{$M$-ary phase shift keying}                       
\acro{M-APSK}[${M}$-PSK]{$M$-ary asymmetric PSK} 
\acro{MTC}{machine type communication}
\acro{MGF}{moment generating function} 
\acro{M-QAM}[$M$-QAM]{$M$-ary quadrature amplitude modulation}
\acro{MRC}{maximal ratio combiner}                  
\acro{maxs}[MSO]{maximum sparsity order}                                      
\acro{M2M}{machine to machine}                                                
\acro{MUI}{multi-user interference}
\acro{mMTC}{massive machine type communications}      
\acro{mm-Wave}{millimeter-wave}
\acro{MP}{mobile phone}
\acro{MPE}{maximum permissible exposure}
\acro{MAC}{media access control}
\acro{NB}{narrowband}
\acro{NBI}{narrowband interference}
\acro{NLA}{nonlinear sparse approximation}
\acro{NLOS}{Non-Line of Sight}
\acro{NTIA}{National Telecommunications and Information Administration}
\acro{NTP}{National Toxicology Program}
\acro{NHS}{National Health Service}
\acro{NB-IoT}{narrowband Internet of things}

\acro{OC}{optimum combining}                             
\acro{OC}{optimum combining}
\acro{ODE}{operational distortion-energy}
\acro{ODR}{operational distortion-rate}
\acro{OFDM}{orthogonal frequency-division multiplexing}
\acro{omp}[OMP]{orthogonal matching pursuit}
\acro{OSMP}[OSMP]{orthogonal subspace matching pursuit}
\acro{OQAM}{offset quadrature amplitude modulation}
\acro{OQPSK}{offset QPSK}
\acro{OFDMA}{Orthogonal Frequency-division Multiple Access}
\acro{OPEX}{Operating Expenditures}
\acro{OQPSK/PM}{OQPSK with phase modulation}
\acro{SS}{{secondary source}}
\acro{SD}{{secondary destination}}

\acro{PAM}{pulse amplitude modulation}
\acro{PAR}{peak-to-average ratio}
\acrodef{pdf}[PDF]{probability density function}                      
\acro{PDF}{probability density function}
\acrodef{p.d.f.}[PDF]{probability distribution function}
\acro{PDP}{power dispersion profile}
\acro{PMF}{probability mass function}                             
\acrodef{p.m.f.}[PMF]{probability mass function}
\acro{PN}{pseudo-noise}
\acro{PPM}{pulse position modulation}
\acro{PRake}{Partial Rake}
\acro{PSD}{power spectral density}
\acro{PSEP}{pairwise synchronization error probability}
\acro{PSK}{phase shift keying}
\acro{PD}{power density}
\acro{8-PSK}[$8$-PSK]{$8$-phase shift keying}
\acro{PR}{primary receiver}
\acro{PTs}{primary transmitters }
 
\acro{FSK}{frequency shift keying}

\acro{QAM}{Quadrature Amplitude Modulation}
\acro{QPSK}{quadrature phase shift keying}
\acro{OQPSK/PM}{OQPSK with phase modulator }

\acro{RD}[RD]{raw data}
\acro{RDL}{"random data limit"}
\acro{ric}[RIC]{restricted isometry constant}
\acro{rict}[RICt]{restricted isometry constant threshold}
\acro{rip}[RIP]{restricted isometry property}
\acro{ROC}{receiver operating characteristic}
\acro{rq}[RQ]{Raleigh quotient}
\acro{RS}[RS]{Reed-Solomon}
\acro{RSC}[RSSC]{RS based source coding}
\acro{r.v.}{random variable}                               
\acro{R.V.}{random vector}
\acro{RMS}{root mean square}
\acro{RFR}{radiofrequency radiation}
\acro{RIS}{reconfigurable intelligent surface}
\acro{RISs}{reconfigurable intelligent surfaces}
\acro{6G}{ sixth generation}

\acro{SA}[SA-Music]{subspace-augmented MUSIC with OSMP}
\acro{SCBSES}[SCBSES]{Source Compression Based Syndrome Encoding Scheme}
\acro{SCM}{sample covariance matrix}
\acro{SEP}{symbol error probability}
\acro{SER}{symbol error rate}
\acro{SG}[SG]{sparse-land Gaussian model}
\acro{SIMO}{single-input multiple-output}
\acro{SINR}{signal-to-interference plus noise ratio}
\acro{SIR}{signal-to-interference ratio}
\acro{SISO}{single-input single-output}
\acro{SMV}{single measurement vector}
\acro{SNR}[\textrm{SNR}]{signal-to-noise ratio} 
\acro{sp}[SP]{subspace pursuit}
\acro{SW}{sync word}
\acro{SAR}{specific absorption rate}
\acro{SSB}{synchronization signal block}
\acro{SR}{secondary receiver}
\acro{ST}{secondary transmitter}

\acro{TH}{time-hopping}
\acro{ToA}{time-of-arrival}
\acro{TR}{transmitted-reference}
\acro{TW}{Tracy-Widom}
\acro{TWDT}{TW Distribution Tail}
\acro{TCM}{trellis coded modulation}
\acro{TDD}{time-division duplexing}
\acro{TDMA}{time division multiple access}

\acro{UAV}{unmanned aerial vehicle}
\acro{uric}[URIC]{upper restricted isometry constant}
\acro{urict}[URICt]{upper restricted isometry constant threshold}
\acro{UWB}{ultrawide band}
\acro{UWBcap}[UWB]{Ultrawide band}   
\acro{URLLC}{Ultra Reliable Low Latency Communications}
         
\acro{wuric}[UWRIC]{upper weak restricted isometry constant}
\acro{wurict}[UWRICt]{UWRIC threshold}                
\acro{UE}{user equipment}
\acro{UL}{uplink}
\acro{URLLC}{ultra reliable low latency communications}

\acro{WiM}[WiM]{weigh-in-motion}
\acro{WLAN}{wireless local area network}
\acro{wm}[WM]{Wishart matrix}                               
\acroplural{wm}[WM]{Wishart matrices}
\acro{WMAN}{wireless metropolitan area network}
\acro{WPAN}{wireless personal area network}
\acro{wric}[WRIC]{weak restricted isometry constant}
\acro{wrict}[WRICt]{weak restricted isometry constant thresholds}
\acro{wrip}[WRIP]{weak restricted isometry property}
\acro{WSN}{wireless sensor network}                        
\acro{WSS}{wide-sense stationary}
\acro{WHO}{World Health Organization}
\acro{Wi-Fi}{wireless fidelity}

\acro{sss}[SpaSoSEnc]{sparse source syndrome encoding}

\acro{VLC}{visible light communication}
\acro{VPN}{virtual private network} 
\acro{RF}{radio frequency}
\acro{FSO}{free space optics}
\acro{IoST}{Internet of space things}

\acro{GSM}{Global System for Mobile Communications}
\acro{2G}{second-generation cellular networks}
\acro{3G}{third-generation cellular networks}
\acro{4G}{fourth-generation cellular networks}
\acro{5G}{5th-generation cellular networks}	
\acro{gNB}{next generation node B base station}
\acro{NR}{New Radio}
\acro{UMTS}{Universal Mobile Telecommunications Service}
\acro{LTE}{Long Term Evolution}

\acro{QoS}{Quality of Service}
\end{acronym}

\def\Pr{\mathrm{Pr}}
\def\Pout{P_{out}}
\def\Pout{\textrm{P}_{out}}
\def\Poutasy{\textrm{P}_{out}^{asy}}
\def\Pe{{\overline{P_e}}}
\def\Peasy{{\overline{P_e}^{asy}}}
\def\Ce{C}
\def\Ceasy{C^{asy}}

\title{Performance Analysis of RIS-Assisted Spectrum Sharing Systems  }

\author{Yazan H. Al-Badarneh,~\IEEEmembership{Member,~IEEE}, Mustafa K. Alshawaqfeh,~\IEEEmembership{Member,~IEEE}, Osamah S. Badarneh,~\IEEEmembership{Member,~IEEE}, Yazid M. Khattabi,~\IEEEmembership{Member,~IEEE}
\thanks{Y. H. Al-Badarneh and Y. M. Khattabi are with the Department of Electrical Engineering, The University of Jordan, Amman, 11942 (email: {yalbadarneh, y.khattabi}@ju.edu.jo).}

\thanks{Y. M. Khattabi is with College of Engineering and Technology, American University of the Middle East, Egaila 54200, Kuwait. e-mail:
yazid.khattabi@aum.edu.kw.}

\thanks{O. S. Badarneh and M. K. Alshawaqfeh are with the Electrical Engineering Department, School of Electrical Engineering and Information Technology, German Jordanian University, Amman 11180, Jordan (e-mail: {Osamah.Badarneh, mustafa.shawaqfeh}@gju.edu.jo).}}
\markboth{}{Al-Badarneh {\MakeLowercase{\textit{et al.}}}: Performance Analysis of RIS-Assisted Spectrum Sharing Systems}
\maketitle

\color{black} 
\begin{abstract} 


We propose a  \ac{RIS}-assisted underlay spectrum sharing system, in which a \ac{RIS}-assisted secondary network shares the spectrum licensed for a primary network. The secondary network consists of a \ac{SS}, an \ac{RIS}, and a \ac{SD}, operating in a Rician fading environment.  We study the performance of the secondary network while considering a peak power constraint at the \ac{SS} and an interference power constraint at the \ac{PR}. Initially, we characterize the statistics of the \ac{SNR} of the \ac{RIS}-assisted secondary network by deriving novel analytical expressions for the \ac{CDF} and \ac{PDF} in terms of the incomplete $H$-function. Building upon the \ac{SNR} statistics, we analyze the outage probability, ergodic capacity, and average bit error rate, subsequently deriving novel exact expressions for these performance measures. Furthermore, we obtain novel asymptotic  expressions for the performance measures of interest when the peak power of the \ac{SS} is high. Finally, we conduct exhaustive Monte-Carlo simulations to confirm the correctness of our theoretical analysis. 

\end{abstract} 

\normalcolor

	\acresetall 
	

\section{Introduction}
Since the advent of its first commercial version in 2018, and as compared to its previous generations, the fifth generation (5G) has shaped a distinguished paradigm of mobile communications with the following emerged applications; namely, enhanced mobile broadband (eMBB), ultrareliable low-latency communications (URLLC) and massive machine-type communications (mMTC) \cite{6736746}, \cite{dang2020should}. However, the ever-increasing current data-hungry applications, such as online gaming, high-definition video streaming, holographic videos, smart homes, mobile shopping, etc., has tremendously augmented the wireless data traffic volume to the extent that 5G would not be capable to support in the coming few years. Motivated by this challenge, industry and academia have recently started conceptualizing the next generation (the\ac{6G})  of wireless communication systems aimed at supporting communication services for future wireless requirements and applications \cite{8808168}.

The \ac{6G}  is expected to provide huge data coverage and to allow tremendous number of subscribers and congested areas of small devices to be connected efficiently and with unconventional high data rate speeds. Nevertheless, attaining all of these \ac{6G} expected requirements and supporting all  its applications, no wonder, requires multiple enabling technologies along with radical solutions to all arising communication engineering challenges, especially, those related to physical layer aspects and fading environments. The random nature of fading environments and the disruptive interactions of the propagated waves with the adjacent objects are the main causes of signals degradation in wireless communication systems. In this context, and within the scope of smart radio environments, there is an  increasing interest in proposing creative communication schemes in which the randomness of the fading environment is to be utilized to improve received signals quality.  A brand-new example is the technology of \ac{RISs}, which have been designed to allow network operators to control reflection, refraction, and scattering characteristics of propagated radio waves in a manner yielding to reducing the negative impacts of fading environments and improving received signals quality \cite{8801961}\nocite{renzo2019smart}-\cite{8466374 }.  \textcolor{white}{ \ac{RIS} }  

The \ac{RISs} are  surfaces implemented from electromagnetic (EM) materials that can be electronically tuned to control the amplitude, phase, frequency, and polarization (wavefront) of propagated signals without the need to complex decoding and encoding processes \cite{8259235 }\nocite{8449754}-\cite{8351817 }. In addition, \ac{RISs} are nearly passive entities, do not require dedicated power sources, can be deployed easily, do not produce noise, and provide full-duplex transmission; traits that are readily essential to support various applications in \ac{6G} networks.  All of these distinguishable aspects have made \ac{RISs} one of the most recent attractive technologies in both academia and industry. In the following, we review the literature of RIS-assisted wireless communications from a performance analysis perspective. 

\subsection{Performance of \ac{RIS}-assisted systems}
In \cite{8796365}, the error performance of \ac{RIS}-assisted communication systems is studied, where the maximized end-to-end (e2e) \ac{SNR} of RIS-assisted networks is assumed to follow a non-central chi-square random variable (RV)  thanks to the central limit theorem (CLT). Based on the \ac{SNR} statistics, the average \ac{SEP} for M-ary \ac{PSK} and square M-ary quadrature amplitude modulation (M-QAM) are investigated. Accurate analytical approximations to the distributions of the received \ac{SNR} for different \ac{RIS}-based wireless system setups are provided in \cite{9144510}. In addition, closed-form and asymptotic expressions for the average channel capacity and the average \ac{BER} are derived. The outage probability and the achievable diversity of RIS-assisted communication systems over generalized fading channels and under the effect of phase noise are investigated in\cite{9738798}. The ergodic capacity and the average \ac{SER} of \ac{RIS}-aided systems over Rayleigh fading channels have been studied in\cite{9095301}. Very recently, a stochastic geometry approach was adopted to assess the coverage probability and average achievable rate of RIS-assisted large-scale networks \cite{10064007}.

RIS-assisted networks over Rician fading channels have gained much attention recently \cite{9146875}\nocite{9247315}-\nocite{9345753} \nocite{9609960}\nocite{9771689 } \cite{9483903}. The Rician model is more general than Rayleigh model and is desirable in realistic communication settings to account for the exitance of \ac{LOS} components between the transmitter and the \ac{RIS}, and between the \ac{RIS} and  the receiver. In \cite{9146875}, the authors analyze the ergodic capacity and the outage probability of \ac{RIS}-aided \ac{SISO} systems over Rician fading channels and in the presence of a direct channel between the transmitter and receiver. In \cite{9247315}, authors study the error performance, data rate of \ac{RIS}-aided systems with single or multiple \ac{RISs} in indoor and outdoor environments. Highly accurate closed-form approximations for the e2e \ac{SNR} statistics of RIS-assisted networks in Rician fading are derived in \cite{9345753}. The authors build upon the obtained statistics to derive closed-form expressions for the ergodic capacity and the average \ac{SER}. The outage performance of \ac{RIS}-aided systems with statistical \ac{CSI} under the Rician fading model is investigated \cite{9609960}. A closed-form expression for the outage probability in terms of Marcum Q-function is derived. Very recently, the authors in \cite{9771689} derive exact analytical expressions for the outage probability, the ergodic capacity, and the average \ac{BER} of \ac{RIS}-aided communications over the sum of cascaded Rician channels. In \cite{9483903}, a robust and secure multiuser \ac{MISO} downlink systems self-sustainable \ac{RIS} was considered. The authors explored the joint optimization of beamformers at an access point (AP) and the phase shifts, along with the energy harvesting schedule at the RIS, aiming to maximize the overall system sum rate.

\subsection{Spectrum sharing systems}

The tremendously increasing number of communicating devices in future \ac{6G} wireless communications will undoubtedly lead to spectrum scarcity, which is a significant problem in modern wireless networks  \cite{9214390}. Spectrum scarcity can be addressed through spectrum sharing-empowered techniques, in which, same frequency band can be shared by multiple communicating devices, but under some interference management mechanisms. In this context, \ac{CR} is defined as an efficient spectrum sharing technique that allows secondary network devices to wisely share same frequency bands licensed for primary network devices \cite{van2017performance}, \cite{8719977}. In \ac{CR} systems, two main spectrum sharing modes can be employed: (i) the overlay mode and (ii) the underlay mode\cite{4840529}. In the overlay mode, secondary devices utilize frequency bands unoccupied by idle primary devices to accomplish their data transmissions. Nevertheless, in the underlay mode, secondary devices perform their data transmissions simultaneously with primary devices, but with limited transmit power such that the interference at primary devices does not exceed a predefined level. 

Envisioning \ac{RIS} technology in underlay \ac{CR} networks has attracted much research interest recently as an efficient solution to improve the performance of such networks. For instance, the \ac{RIS} has been considered as an indispensable approach to further enhance the energy and spectrum efficiency in \ac{CR} networks \cite{9146170}, \cite{9235486}. In addition, the deployment of the \ac{RIS} in \ac{CR} networks can significantly enhance the secrecy outage probability and the achievable secrecy rate \cite{9562609}, \cite{9402750}. In what follows, we provide a comprehensive literature review on \ac{RIS}-aided \ac{CR} systems.

The \ac{RIS} technology is employed to aid  \ac{MIMO} \ac{DL} communication in \ac{CR} systems \cite{9146170}, where the achievable weighted sum rate of the secondary users is maximized under a transmit power limit at the \ac{ST}, interference temperature constraints on the primary users (PUs), and unit modulus constraints. In \cite{9235486}, the authors propose multiple RISs to assist the \ac{DL} communication of \ac{MISO} CR systems. The achievable rate of \ac{SU} is maximized subject to a transmit power limit at the \ac{ST} and interference temperature constraints on the PUs. In \cite{9183907}, the authors consider a resource allocation problem for multiuser \ac{FD} CR systems in the presence of a RIS. The study aimed to maximize the sum rate by collectively optimizing the \ac{DL} transmit beamforming and \ac{UL} receive beamforming vectors, the transmit power of the \ac{UL} users, and the phase shift matrix at the \ac{RIS}. In \cite{9419976}, the authors suggest and examine two separate configurations of \ac{RIS} deployment to improve the performance of \ac{CR} networks. Analytical expressions for the false alarm and detection probabilities of the two \ac{RIS} configurations are obtained. In \cite{9521976}, a robust beamforming design is proposed for \ac{RIS}-aided CR systems. The authors investigated the beamforming design considering two types of error models for PU-related channels: (i) bounded \ac{CSI} and (ii) statistical CSI..

\subsection{  Contributions } 

In the context of performance analysis, most of the aforementioned works consider the performance of \ac{RIS}-assisted traditional communication systems with no spectrum sharing, where the Rayleigh and Rician fading models are considered. In the context of \ac{RIS}-aided spectrum sharing  systems, all previous works focus on resource allocation problems to optimize the performance of the secondary network in terms of data rate. The deployment of the \ac{RIS} in spectrum sharing  systems can substantially improve the performance with regard to the outage probability, ergodic capacity, and average \ac{BER}.

Only a few studies have recently focused on analyzing the performance in terms of the outage probability of \ac{RIS}-aided spectrum sharing  systems. 
For instance, the outage probability analysis of a simultaneously transmitting and reflecting reconfigurable intelligent surface (STAR-RIS)-empowered cognitive non-terrestrial vehicle network with non-orthogonal multiple access (NOMA) was considered in \cite{10093979}. The outage probability of a RIS-aided cognitive radio network where the RIS is used only to eliminate the interference from the secondary transmitter to the primary receiver was studied in \cite{9773290}.  Very recently,  the characteristics of the  signal-to-Interference-plus-Noise Ratio (SINR) of a RIS-aided device-to-device (D2D) communication system operating in underlay mode was considered in \cite{10095527}. The authors therein considered a novel SINR maximization problem that jointly optimizes transmit power budget at the D2D source and the RIS deployment.

\color{black} 

In this work, we present a novel \ac{RIS}-assisted underlay spectrum sharing system, in which a secondary network assisted by a RIS shares the spectrum dedicated to a primary network. The secondary network comprises a \ac{SS}, a \ac{RIS}, and a \ac{SD}, operating in a Rician fading environment.  This work aims to provide a comprehensive  performance of the envisioned \ac{RIS}-assisted secondary network. To the best of the authors' knowledge, such  a comprehensive performance analysis has not been addressed in the existing literature.  The advantages of the proposed system lie in the following aspects: 

\begin{itemize}

\item The \ac{RIS}-assisted secondary network shares the spectrum of the primary network without deteriorating the performance of the primary link. This can be ensured by adopting transmit power adaptation at the \ac{SS}. In particular, the \ac{SS} can adjust its transmit power appropriately to limit the interference power at the \ac{PR} caused by the \ac{SS} to below a specified threshold.  Additionally, the transmit power of the \ac{SS} is subject to a peak power constraint $P$.

\normalcolor 
\item The deployment of the \ac{RIS} in the secondary network is of significant practical importance due to the following reasons: 
\begin{itemize}
\item In a situation where no direct link is available between the \ac{SS} and the \ac{SD}, the \ac{RIS} is deployed in an effort to make the communication between the \ac{SS} and the \ac{SD} not only possible but also reliable. 

\item Unlike relaying \cite{9095301} and backscatter \cite{9455142} communication techniques, the reflected wave from the \ac{RIS} can be effectively controlled and programmed in real-time due to the adjustable phase shifts of the \ac{RIS}. The RIS is mainly deployed for receiving a signal from the \ac{SS} and focusing the reflected signal towards the \ac{SD} exclusively. As a result, any interference signal reflected from the \ac{RIS} towards the \ac{PR} is disregarded due to the limited coverage area of the \ac{RIS}, which restricts the propagation of interference towards the \ac{PR} \cite{9773290}, \cite{10095527} and \cite{9000593}.

\end{itemize}

 \item The \ac{RIS}-assisted secondary network  is assumed to undergo Rician fading. The Rician model is more general than Rayleigh model and is desirable in realistic communication settings to account for the existence of \ac{LOS} components between the \ac{SS} and \ac{RIS}, and between the \ac{RIS} and the \ac{SD} \cite{9345753}. The widely adopted Rayleigh fading model is a special case of the Rician model when the Rician $K$-factor equals zero.

\end{itemize}

\color{black} 
It is worth emphasizing that expressing the \ac{CDF} and \ac{PDF} of the \ac{SNR} in the \ac{RIS}-assisted secondary network, while considering both the Rician model and the transmit power adaption at the \ac{SS}, in closed-form is very challenging. Consequently, the performance analysis is mathematically intricate, requiring the solution of difficult integrals to quantify the ergodic capacity and the average \ac{BER}. Such integrals are tackled using the theory of the incomplete $H$-function. To our best knowledge, this work represents the first application of the incomplete $H$-function in analyzing wireless communication systems. The key contributions of this work can be outlined as follows: 

\begin{itemize} 

\item In terms of an incomplete $H$-function, we develop novel analytical expressions for the \ac{CDF} and \ac{PDF} of the \ac{SNR}.
The outage probability can be evaluated immediately afterward using the provided \ac{CDF}. 
\item Building upon the obtained \ac{PDF} and \ac{CDF}, we provide the cumbersome ergodic capacity and average \ac{BER} analysis. Novel analytical expressions in terms  of the incomplete $H$-function and the $H$-function are obtained. 

\item Utilizing the asymptotic properties of the incomplete $H$-function, we obtain accurate asymptotic expressions for the outage probability, ergodic capacity, and the average \ac{BER} when the peak power of the \ac{SS} is high.

\end{itemize}

\normalcolor

\medskip

The subsequent content of this work is organized as follows. The system model is articulated in Section II. The exact and the asymptotic performance of the \ac{RIS}-assisted secondary network are investigated in Section III and Section IV, respectively. Section V is dedicated  to  investigating the numerical results. Finally, the conclusions are drawn in Section VI.  

\color{black} 
\section{System model}
\begin{figure*}[h]\label{fig:01}
\begin{center}
\includegraphics[width=1\textwidth]{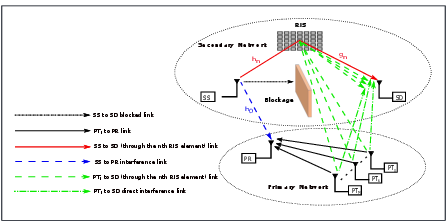}
\caption{A \ac{RIS}-aided secondary network shares the spectrum allocated to a primary network consisting of $M$  \ac{PTs} simultaneously communicating with a \ac{PR}.}
\end{center}
\end{figure*} 

We take into consideration an underlay spectrum sharing system, as shown in Fig. 1, where a secondary network shares the spectrum allotted to a primary network comprising  $M$ \ac{PTs} simultaneously communicating with a \ac{PR}.  The  \ac{PTs} and the \ac{PR} are single-antenna devices. The secondary network is a \ac{RIS}-assisted environment consisting of a \ac{SS}, an  \ac{RIS} of $N$ reflecting elements, and a \ac{SD}. The  \ac{SS} and  the \ac{SD} are single-antenna terminals.  We use $h_{0}$ to denote the channel coefficient from  the \ac{SS} to the \ac{PR}. For the secondary network, we denote the channel coefficients from the \ac{SS} to the $n^{th}$ reflecting element and from the $n^{th}$ reflecting element to the  \ac{SD} by  $h_{n}$ and $g_{n}$, respectively. The channel coefficients of the secondary network can be expressed as $h_{n}=\alpha_{n} e^{-j\theta_{n}}$ and $g_{n}=\beta_{n} e^{-j\psi_{n}}$, where $\alpha_{n}$ and $\beta_{n}$ represent the channel envelopes, and $\theta_{n}$ and $\psi_{n}$ represent the phase shifts. The envelopes $\alpha_{n}$ and $\beta_{n}$ are independent random variables (RVs), for $n=1, 2, ...., N$.

We assume that the \ac{SS} will communicate with the \ac{SD} through the \ac{RIS}, without a direct link between the \ac{SS} and the \ac{SD} due to presence of natural or man-made obstacles \cite{8796365}, \cite{9345753} and \cite{9138463}. Additionally, we take into account the existence of a \ac{LOS} component between the \ac{SS} and the \ac{RIS}, and between the \ac{RIS} and the \ac{SD}. Accordingly, $\alpha_{n}$, $n=1, 2, ...., N$ can be modeled as independent identically distributed (i.i.d) Rician RVs with shape parameter $K_{1}$ and scale parameter $\Omega_{1}$. Similarly, $\beta_{n}$, $n=1, 2, ...., N$ can be also viewed as i.i.d Rician RVs with shape parameter $K_{2}$ and scale parameter $\Omega_{2}$  \footnote{The assumption of i.i.d RIS channels has been widely adopted in the literature  since it facilitates the development of mathematically tractable formulations and serves as a foundational framework for understanding the performance. The consideration of correlated RIS Rician channels requires the development of different mathematical formulations, which remains as a future research endeavor.}.  The shape parameter and scale parameter can be, respectively, characterized  as $K_{j}=\frac{\mu^{2}_{j}}{2 \sigma^{2}_{j}} $ and $\Omega_{j}=\mu^{2}_{j}+2 \sigma^{2}_{j}$, $j \in \{1,2\}$, where $\mu^{2}_{j}$ denotes the power in \ac{LOS} component and $2 \sigma^{2}_{j}$ denotes the power in the \ac{NLOS} component.

To this end, we assume that the primary and secondary networks are distant from each other. Accordingly, $|h_{0}|$ can be presumed to undergo Rayleigh fading implying that $|h_{0}|^{2}$ is an exponential RV with mean $\lambda$. Furthermore, we assume that the interference caused by the \ac{PTs} to the \ac{SD} (i.e., direct \ac{PTs} $\to$ \ac{SD} and \ac{PTs} $\to$ \ac{RIS} $\to$ \ac{SD}) is translated into white Gaussian noise at the \ac{SD}. This assumption is justified by the central limit theorem (CLT), which indicates that the interference from multiple \ac{PTs} converges to a Gaussian distribution due to the presence of multiple \ac{PTs} and multiple \ac{RIS} elements \cite{4786488}. Accordingly, the interference-plus-noise at the \ac{SD} can be effectively modeled as an additive white Gaussian noise (AWGN) with a zero mean and variance of $N_{0}$.

As shown in \cite{8796365}, the  instantaneous \ac{SNR} at the \ac{SD} is maximized by eliminating the channel phase shifts $\psi_{n}$ and $\theta_{n}$. Hence, the maximized  instantaneous \ac{SNR} at the \ac{SD}, denoted by $\rho$, is given by  \footnote{Similar to\cite{8796365},  \cite{9144510} and \cite{9345753},  perfect knowledge of the \ac{RIS} channels CSI is assumed to be available at the \ac{SS}. Furthermore, perfect CSI can be acquired at the transmitter- or the receiver-side \cite{9366805},  \cite{9527799}. The impact of imperfect CSI is beyond the scope of this work and is left as a future research endeavor.}
\begin{equation}\label{eq:L1}
\rho= \frac{ P_{s} \left( \sum_{n=1}^{N} \alpha_{n}\, \beta_{n} \right)^{2}}{N_{0}},
\end{equation}
where $P_{s}$ is the transmit power of the \ac{SS}.

In spectrum sharing systems, the \ac{PR} mandates that the instantaneous interference caused by the  \ac{SS} is below predetermined threshold $Q$, which indicates the interference power limit at the \ac{PR}  \footnote{ The interference signal reflected from the \ac{RIS} towards the \ac{PR} is disregarded due to the limited coverage area of the \ac{RIS}, which restricts the propagation of interference towards the \ac{PR} \cite{9773290}, \cite{10095527} and \cite{9000593}. }. Meanwhile, the transmit power of the \ac{SS} is subject to a peak power constraint $P$.  Accordingly, the transmit power at the \ac{SS} is given by 
 \begin{equation}\label{eq:L2}
 P_{s} =\min \left( \frac{Q}{|h_{0}|^{2}}, P \right),
\end{equation} 
where $P$ is the peak power of the \ac{SS}.   Capitalizing on (\ref{eq:L2}), $\rho$ in  (\ref{eq:L1}) can be rewritten as 
\begin{equation}\label{eq:L3}
\begin{split}
\rho &= \min \left( \frac{Q}{|h_{0}|^{2}}, P \right)  \frac{R^{2}}{N_{0}} \\
 &= 
\begin{cases}
    P R^{2} , & |h_{0}|^{2} \leq \frac{Q}{P} \\
    \frac{Q R^{2}}{ |h_{0}|^{2}}, & |h_{0}|^{2} > \frac{Q}{P}
\end{cases},
\end{split}
\end{equation}
where $R^{2}=\left( \sum_{n=1}^{N} \alpha_{n}\, \beta_{n} \right)^{2}$, and $N_{0}$ is set to unity without loss of generality. 

To this end, the \ac{CDF} of $R^{2}$ is given by \cite[Theorem~1]{9345753}
\begin{equation}\label{eq:L4}	
F_{R^{2}}(z)= \frac{\gamma \left(v,\frac{\sqrt{z}}{b}\right)}{\Gamma(v)}, \ \ z \geq 0, 
\end{equation}
where $\gamma(\cdot,\cdot)$ is the lower incomplete gamma function \cite[Eq. (8.350.1)]{jeffrey2007table}.  If we define $R_{n}= \alpha_{n}\, \beta_{n}$, where $R_{n}$ are i.i.d RVs, $n=1, 2, ...., N$, then the parameters $v$ and $b$ are dependent on the mean and variance of $R_{n}$, and can be respectively expressed as  \cite{9345753}
\begin{equation*}	
v=\frac{N \left( E\left[ R_{n} \right] \right)^{2}}{\text{var}\left(R_{n}\right)}, \ \ \ \ \ 
b=\frac{\text{var}\left(R_{n}\right)}{ E\left[R_{n}\right]}, 
\end{equation*}
where  $E\left[ R_{n} \right]$ and $\text{var}\left(R_{n}\right)$ are obtained as in \cite[Eq. (4)]{9345753} and \cite[Eq. (5)]{9345753}, respectively. 

\section{ Exact Performance Analysis}
This section provides the performance of the \ac{RIS}-assisted secondary network. We start with deriving analytical expressions for the \ac{CDF} and \ac{PDF} of the maximized instantaneous \ac{SNR} $\rho$ in (\ref{eq:L3}). The derived expressions are to be utilized in the sequel to analyze the outage performance, ergodic capacity and average \ac{BER} performance. 
\subsection{SNR Statistics} 

In Theorem 1 below, the \ac{CDF} and \ac{PDF} of $\rho$ are provided in terms of incomplete $H$-functions. 

\noindent{\textbf{Theorem 1:}} The \ac{CDF} and \ac{PDF} of $\rho$ are, respectively, given by  
\begin{equation}\label{eq:L5}
	\begin{split}
F_{\rho} (z)&= \frac{\gamma \left(v,  \sqrt{\frac{z}{b^2 P}} \, \right)}{\Gamma(v)}  \left( 1- e^{-\frac{Q}{P \lambda}} \right)   \\
&\quad+  \frac{1}{\Gamma(v)} \Gamma_{2,2}^{1,2}\left[ \left.  \sqrt{\frac{z\lambda}{b^2 Q}} \right| \begin{smallmatrix} (0, \frac{1}{2}, \frac{Q}{P \lambda} ), (1, 1) \\ ~\\ (v, 1), (0, 1) \end{smallmatrix} \right], 
	\end{split}
\end{equation}
and 
\begin{equation} \label{eq:L6}  
\begin{split}  
f_{\rho} (z)&= \frac{z^{\frac{v}{2}-1}       e^{- \sqrt{\frac{z}{b^2 P}}}     }{2(b^{2} P)^{\frac{v}{2}} \Gamma(v)}  \left( 1- e^{-\frac{Q}{P \lambda}} \right)   \\
&\quad+  \frac{z^{-1}}{\Gamma(v)} \Gamma_{3,3}^{1,3}\left[ \left.  \sqrt{\frac{z\lambda}{b^2 Q}} \right| \begin{smallmatrix} (0,\frac{1}{2}, \frac{Q}{P \lambda} ), (0, \frac{1}{2}), (1, 1) \\ ~\\ (v, 1), (0, 1), (1, \frac{1}{2}) \end{smallmatrix} \right], 
\end{split}
\end{equation}
where $\Gamma_{p,q} ^{m,n}(\cdot )$ is the incomplete $H$-function \cite[ Eq. (2.3)] {srivastava2018incomplete} and $\Gamma(\cdot)$ is the gamma function  \cite[Eq. (8.310.1)]{gradshteyn2014table}.

\noindent \textit{Proof:} See Appendix A. 

\subsection{Outage Probability}
A fundamental measure of the communication link reliability is the outage probability, which is defined as the probability that the received \ac{SNR} falls below a predefined threshold $\gamma_{\mathrm{th}}$. Correspondingly, the outage probability of the secondary network, denoted by $\Pout(\gamma_{\mathrm{th}})$, is given by 
 \begin{equation}\label{eq:L7}
\begin{split}
\Pout(\gamma_{\mathrm{th}}) &= F_{\rho}(\gamma_{\mathrm{th}})\\
&= \frac{\gamma \left(v,  \sqrt{\frac{\gamma_{\mathrm{th}}}{b^2 P}} \, \right)}{\Gamma(v)}  \left( 1- e^{-\frac{Q}{P \lambda}} \right)   \\
&\quad+  \frac{1}{\Gamma(v)} \Gamma_{2,2}^{1,2}\left[ \left.  \sqrt{\frac{\gamma_{\mathrm{th}} \lambda}{b^2 Q}} \right| \begin{smallmatrix} (0,\frac{1}{2}, \frac{Q}{P \lambda} ), (1,1) \\ ~\\ (v, 1), (0,1) \end{smallmatrix} \right]. 
\end{split} 
\end{equation}

\subsection{Ergodic Capacity}
We now provide the ergodic capacity analysis of the secondary network under consideration.  The instantaneous capacity of the secondary network for a given realization of the RV $\rho$, denoted by $C_{\rho}$, is given by  \cite{proakis2008digital}
\begin{equation}\label{eq:L8}
C_{\rho}=\log_{2} (1+\rho)  \quad [\text{bits/s/Hz}]. 
  \end{equation}
In what follows, we use  $C$ to denote the ergodic capacity, which can be obtained by averaging over all realizations of the RV $\rho$ as 
\begin{equation}\label{eq:L9}
\begin{split}
\Ce=& E \left[ \log_{2} \left(1+\rho \right) \right]\\
=& \frac{1}{\ln(2)} \int_{0}^{\infty}  \ln \left( 1+ z \right) f_{\rho} (z) \mathrm{d}z ,\\
\end{split}
\end{equation}
where $f_{\rho} (z)$ is as given in  (\ref{eq:L6}).

In Theorem 2 below, the ergodic capacity is derived in terms of the $H$-function and the incomplete $H$-function. 

\noindent{\textbf{Theorem 2:}} The ergodic capacity of the secondary network is given by  
\begin{equation}\label{eq:L10}
\Ce= \Ce^{(1)}+ \Ce^{(2)}, 
\end{equation}
where $\Ce^{(1)}$ and $\Ce^{(2)}$ are, respectively,  given by 
\begin{equation} \label{eq:L11}  
\Ce^{(1)}=  \frac{\left( 1- e^{-\frac{Q}{P \lambda}} \right)}{\Gamma(v) \ln(2)} H_{3,2}^{1,3}\left[ \left.  {b^2 P} \vphantom{  \sqrt{\frac{\lambda}{b^2 Q}}} \right| \begin{smallmatrix} (1-v, 2) , (1, 1), (1,1) \\ ~\\ (1, 1), (0, 1) \end{smallmatrix} \right], 
\end{equation}

\begin{equation} \label{eq:L12}  
\Ce^{(2)}=  \frac{1}{\ln(2) \Gamma(v)} \Gamma_{3,3}^{2,3}\left[ \left.  \sqrt{\frac{\lambda}{b^2 Q}} \right| \begin{smallmatrix} (0, \frac{1}{2}, \frac{Q}{P \lambda} ), (1, \frac{1}{2}), (1, 1) \\ ~\\ (v, 1), (1, \frac{1}{2}), (0, 1) \end{smallmatrix} \right], 
\end{equation}
where  $H_{p,q} ^{m,n}[\cdot ]$ is the univariate $H$-function \cite[Eq. (1.2)]{mathai2009h}. 

\noindent \textit{Proof:} See Appendix B.

\subsection{Average BER}
Here, we study the average \ac{BER} of the secondary network. We focus on various binary modulation formats with conditional \ac{BER}, denoted by $ P_{e}(\rho) $, is given by \cite{5957242}
\begin{equation}\label{eq:L13}
P_{e}(\rho) = \frac{\Gamma(\zeta, \delta \rho)}{2 \Gamma(\zeta)}, 
\end{equation}
where  $\rho$ is the instantaneous \ac{SNR} in (\ref{eq:L3}), and $\Gamma(\cdot,\cdot )$ is the upper incomplete gamma function \cite[Eq. (8.350.2)]{jeffrey2007table}, and the parameters $\delta$ and $\zeta$ are modulation-specific constants. The average \ac{BER}, denoted by $P_{e}$, can be derived as   \cite{5957242}
\begin{equation}\label{eq:L14}
		P_{e} =\frac{\delta^{\zeta}}{2 \Gamma(\zeta)} \int_{0}^{\infty} e^{-\delta z} z^{\zeta-1} F_{\rho}(z)  \mathrm{d}z.   
\end{equation}

In Theorem 3 below, we obtain the average \ac{BER} of the secondary netowrk in terms of the $H$-function and the incomplete $H$-function.

\noindent{\textbf{Theorem 3:}} The average \ac{BER} of the secondary network is given by  
\begin{equation}\label{eq:L15}
P_{e}= P^{(1)}_{e}+ P^{(2)}_{e}, 
\end{equation}
where $ P^{(1)}_{e}$ and $ P^{(2)}_{e}$ are, respectively,  given by 
\begin{equation} \label{eq:L16}  
 P^{(1)}_{e}=  \frac{\left( 1- e^{-\frac{Q}{P \lambda}} \right)}{2 \Gamma(v) \Gamma(\zeta)} H_{2,2}^{1,2}\left[ \left.   \sqrt{\frac{1}{\delta b^2 P}}  \right| \begin{smallmatrix} (1-\zeta, \frac{1}{2}) , (1, 1) \\ ~\\ (v, 1), (0, 1) \end{smallmatrix} \right], 
\end{equation}

\begin{equation} \label{eq:L17}  
P^{(2)}_{e}=  \frac{1}{2 \Gamma(v) \Gamma(\zeta)} \Gamma_{3,2}^{1,3}\left[ \left.  \sqrt{\frac{\lambda}{\delta b^2 Q}} \right| \begin{smallmatrix} (0, \frac{1}{2}, \frac{Q}{P \lambda} ), (1-\zeta, \frac{1}{2}), (1, 1) \\ ~\\ (v, 1),  (0, 1) \end{smallmatrix} \right].  
\end{equation}

\noindent \textit{Proof:} See Appendix C.

\section{Asymptotic Performance Analysis}
This section provides the asymptotic performance of the secondary network at high values of peak power $P$.  While the exact results obtained in the previous section are general enough to cover various arbitrary values of system parameters, the results of this section are of interest when $P \lambda \gg Q$.  In what follows, we derive asymptotic expressions for the outage probability, ergodic capacity, and average \ac{BER} when $P \lambda \gg Q$.

\subsection{Asymptotic Outage Probability}
Considering $\Pout(\gamma_{\mathrm{th}})$ in (\ref{eq:L7}), the first term will approach zero when $P \lambda \gg Q$ (i.e.,  $( 1- e^{-\frac{Q}{P \lambda}}) \to 0)$. Hence, the asymptotic outage probability, denoted by $\Poutasy(\gamma_{\mathrm{th}})$, can be approximated as 
\begin{equation}\label{eq:L18}
	 \Poutasy(\gamma_{\mathrm{th}}) \approx  \frac{1}{\Gamma(v)} \Gamma_{2,2}^{1,2}\left[ \left.  \sqrt{\frac{\gamma_{\mathrm{th}} \lambda}{b^2 Q}} \right| \begin{smallmatrix} (0,\frac{1}{2}, \frac{Q}{P \lambda} ), (1,1) \\ ~\\ (v, 1), (0,1) \end{smallmatrix} \right]. 
\end{equation}

If we further consider the unlimited peak power scenario (i.e.,  $ P \to \infty$), then  $\Poutasy(\gamma_{\mathrm{th}})$ can be obtained as 
 \begin{equation}\label{eq:L19}
\Poutasy(\gamma_{\mathrm{th}}) \approx  \frac{1}{\Gamma(v)} \Gamma_{2,2}^{1,2}\left[ \left.  \sqrt{\frac{\gamma_{\mathrm{th}} \lambda}{b^2 Q}} \right| \begin{smallmatrix} (0,\frac{1}{2}, 0), (1,1) \\ ~\\ (v, 1), (0,1) \end{smallmatrix} \right]. 
\end{equation}
 In view of \cite[case (4)]{meena2020some}, the incomplete $H$-function in (\ref{eq:L19}) reduces to an $H$-function as 
 \begin{equation}\label{eq:L20}
\Poutasy(\gamma_{\mathrm{th}}) \approx  \frac{1}{\Gamma(v)} H_{2,2}^{1,2}\left[ \left.  \sqrt{\frac{\gamma_{\mathrm{th}} \lambda}{b^2 Q}} \right| \begin{smallmatrix} (0,\frac{1}{2}), (1,1) \\ ~\\ (v, 1), (0,1) \end{smallmatrix} \right]. 
\end{equation}

\subsection{Asymptotic Capacity}
When $P \lambda \gg Q$, the asymptotic ergodic capacity, denoted by  $\Ceasy$, can be approximated by $\Ce^{(2)}$ in  (\ref{eq:L12}).  The term $ \Ce^{(1)}$ in (\ref{eq:L11})  becomes negligible due to $( 1- e^{-\frac{Q}{P \lambda}}) \to 0$.  Accordingly, 
\begin{equation}\label{eq:L21}
\Ceasy \approx  \frac{1}{\ln(2) \Gamma(v)} \Gamma_{3,3}^{2,3}\left[ \left.  \sqrt{\frac{\lambda}{b^2 Q}} \right| \begin{smallmatrix} (0, \frac{1}{2}, \frac{Q}{P \lambda} ), (1, \frac{1}{2}), (1, 1) \\ ~\\ (v, 1), (1, \frac{1}{2}), (0, 1) \end{smallmatrix} \right]. 
\end{equation}

If we further consider the unlimited peak power scenario (i.e.,  $ P \to \infty$), then
\begin{equation}\label{eq:L22}
\begin{split}
\Ceasy & \approx \frac{1}{\ln(2) \Gamma(v)} \Gamma_{3,3}^{2,3}\left[ \left.  \sqrt{\frac{\lambda}{b^2 Q}} \right| \begin{smallmatrix} (0, \frac{1}{2}, 0 ), (1, \frac{1}{2}), (1, 1) \\ ~\\ (v, 1), (1, \frac{1}{2}), (0, 1) \end{smallmatrix} \right]\\ 
&=\frac{1}{\ln(2) \Gamma(v)} H_{3,3}^{2,3}\left[ \left.  \sqrt{\frac{\lambda}{b^2 Q}} \right| \begin{smallmatrix} (0, \frac{1}{2}), (1, \frac{1}{2}), (1, 1) \\ ~\\ (v, 1), (1, \frac{1}{2}), (0, 1) \end{smallmatrix} \right].  
\end{split}
\end{equation}

\subsection{Asymptotic Average BER}
Following the same procedure for deriving the $\Poutasy$, and $\Ceasy$,  the asymptotic BER, denoted by $P^{asy}_{e}$, can be obtained as 
\begin{equation}\label{eq:L23}
		P^{asy}_{e} \approx  \frac{1}{2 \Gamma(v) \Gamma(\zeta)} \Gamma_{3,2}^{1,3}\left[ \left.  \sqrt{\frac{\lambda}{\delta b^2 Q}} \right| \begin{smallmatrix} (0, \frac{1}{2}, \frac{Q}{P \lambda} ), (1-\zeta, \frac{1}{2}), (1, 1) \\ ~\\ (v, 1),  (0, 1) \end{smallmatrix} \right].  
\end{equation}
In addition, as $P \to \infty$, we have
\begin{equation}\label{eq:L24}
		P^{asy}_{e} \approx  \frac{1}{2 \Gamma(v) \Gamma(\zeta)} H_{3,2}^{1,3}\left[ \left.  \sqrt{\frac{\lambda}{\delta b^2 Q}} \right| \begin{smallmatrix} (0, \frac{1}{2} ), (1-\zeta, \frac{1}{2}), (1, 1) \\ ~\\ (v, 1),  (0, 1) \end{smallmatrix} \right]. 
\end{equation}

\normalcolor

\section{ Numerical Results}

In this section, we utilize the derived analytical expressions for the outage probability, the ergodic capacity, and the average BER to investigate numerically the performance of the \ac{RIS}-assisted secondary network for various system parameters. The correctness of analytical expressions is validated by using extensive Monte-Carlo simulations \footnote{The  incomplete $H$-function is not a built-in function in well-known software packages such as \textsc{Matlab} and \textsc{Mathematica}. Hence, we provide an efficient \textsc{Matlab} code for evaluating the incomplete H-function at: https://www.mathworks.com/matlabcentral/fileexchange/161176-incomplete-fox-h-function. }. 

\begin{figure}[h]
\centering
\includegraphics[width =1 \columnwidth]{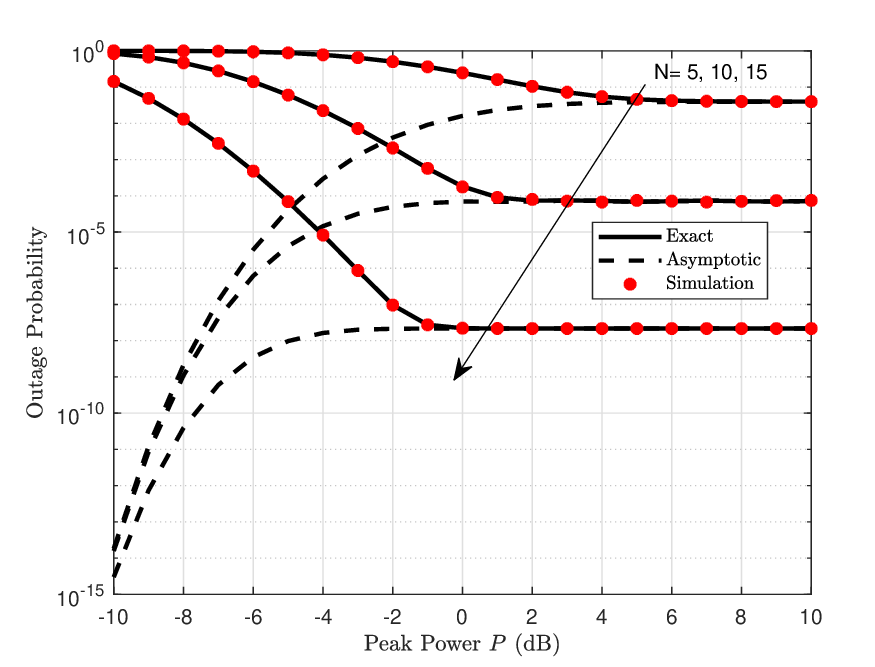}
\caption{Outage probability of the secondary network versus  the \ac{SS} peak power $P$  for various values of RIS elements $N$ at $\lambda= -5 \ \text{dB}$, $\gamma_{\mathrm{th}}= 10 \ \text{dB}$, and $Q=  0 \ \text{dB}$.}
\label{fig_OP_Omega_N}
\end{figure}

It is worth mentioning that our analytical results hold true for arbitrarily chosen parameters. However, for the purpose of comparing the analytical results with the simulation results, we specifically consider the following settings, unless stated otherwise.  In our simulations, we consider $10^{7}$ independent samples (realizations) for performance evaluation. For the channel parameters of the secondary network, we take into consideration a Rician fading environment with the following parameters as in \cite{9345753} : $v_{1} = v_{2} = 1$, $\sigma^{2}_{1} = \sigma^{2}_{2} = 0.5$ (i.e., the corresponding Rician factors are $K_{1} = K_{2} = 1$). In all figures, the analytical, asymptotic, and Monte-Carlo simulation results are represented by solid lines,  dashed lines, and markers only, respectively. 

\color{black}

In Fig. 2, the outage probability of the secondary network is depicted against the \ac{SS} peak power $P$ for different values of \ac{RIS} elements $N$. We first note that the exact outage probability in (\ref{eq:L7}) precisely matches the Monte-Carlo simulations. Additionally, we observe that the asymptotic curves for the outage probability tend to converge to the exact curves as $P \lambda \gg Q$. As expected, the outage performance improves with increasing $N$. It is also evident that the outage performance exhibits saturation due to the peak power constraint at the \ac{SS}.



Fig. 3 depicts the ergodic capacity versus the \ac{SS} peak power $P$ for different values of $N$. We note that the exact ergodic capacity in (\ref{eq:L10}) matches perfectly with the Monte-Carlo simulations. Additionally, it can be observed that the asymptotic expression for ergodic capacity is highly accurate when $P \lambda \gg Q$. From this figure, it is evident that the ergodic capacity improves with increasing $N$ or $P$, but it saturates as $P$ grows large.


\begin{figure}[h]
	\centering
	\includegraphics[width = \columnwidth]{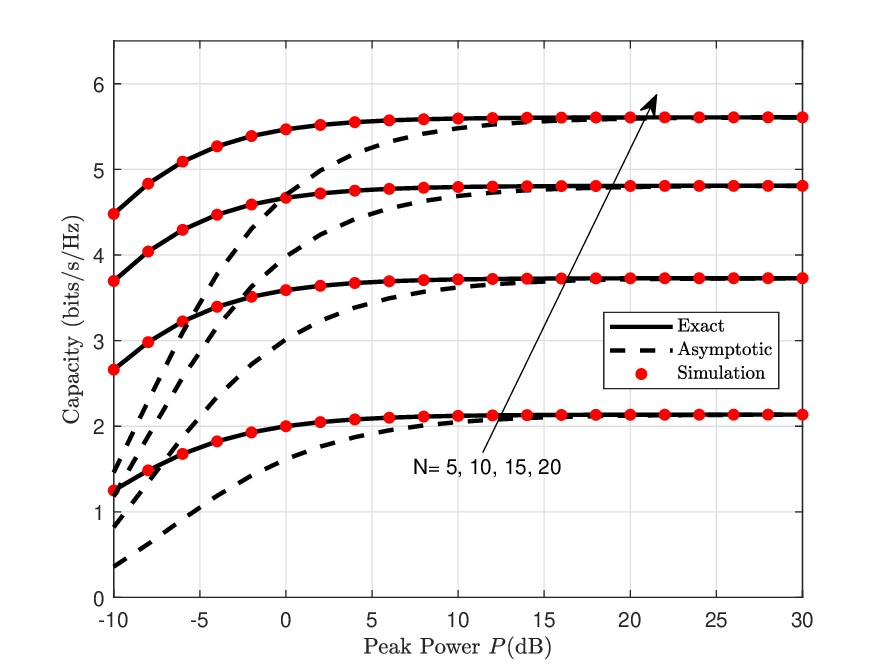}
	\caption{Ergodic Capacity of the secondary network versus  the \ac{SS} peak power $P$  for various values of RIS elements $N$ at $\lambda= 0 \ \text{dB}$ and $Q=  -10 \ \text{dB}$.}
	\label{fig_C_Omega_N}
\end{figure}

\begin{figure}[h]
	\centering
	\includegraphics[width = \columnwidth]{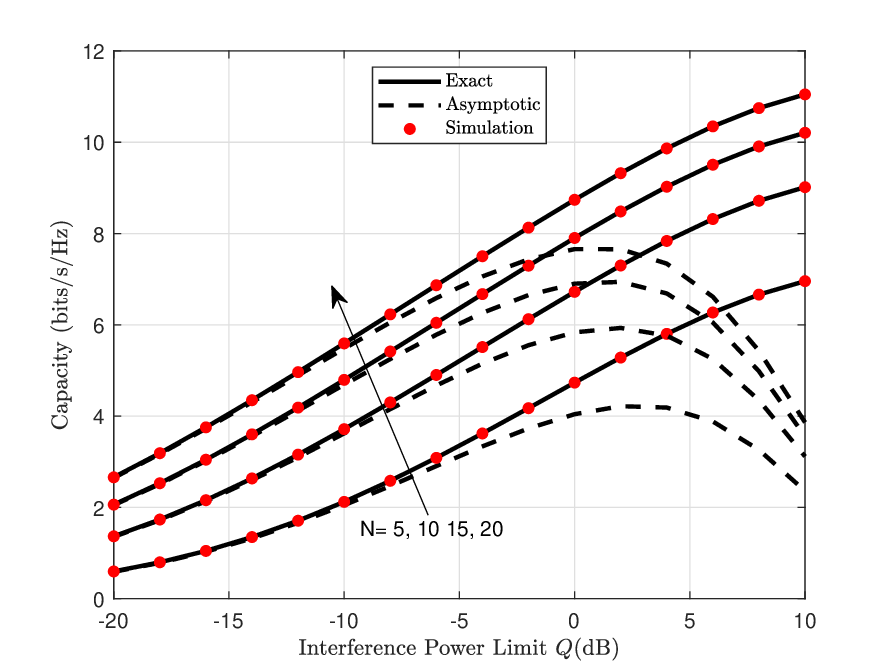}
	\caption{Ergodic Capacity of the secondary network versus the interference power limit  $Q$ for  various values of RIS elements $N$ at  $\lambda= 0 \ \text{dB}$ and $P= 10 \ \text{dB}$.} 
	\label{fig_C_N_Omega}
\end{figure}

Fig. 4 illustrates the ergodic capacity versus the interference power limit $Q$ at the \ac{PR} for different values of $N$. We observe that the exact ergodic capacity improves with increasing $N$, as expected. It is also evident that as the \ac{PR} tolerates higher interference power (i.e., $Q$ increases), the capacity improves. This is attributed to the fact that the \ac{SS} transmit power increases with $Q$ as shown in (\ref{eq:L3}), leading to better performance. Additionally, we note that while the asymptotic ergodic capacity is highly accurate when $P \lambda \gg Q$, it diverges  from the exact capacity as $Q$ increases.


%
%
%
%
%

%

\begin{figure}[h]
	\centering
	\includegraphics[width = \columnwidth]{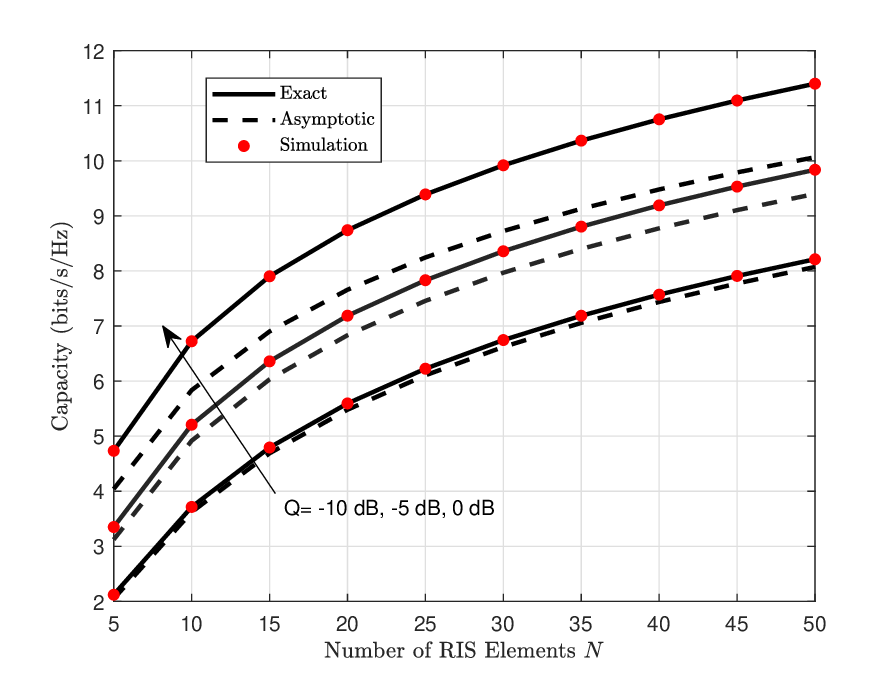}
	\caption{Ergodic Capacity of the secondary network versus the number of RIS elements $N$  for various values of $Q$ at   $\lambda= 0 \ \text{dB}$ and $P= 10 \ \text{dB}$.}
	\label{fig_C_K_Omega}
\end{figure}

\begin{figure}[h]
	\centering
	\includegraphics[width =1 \columnwidth]{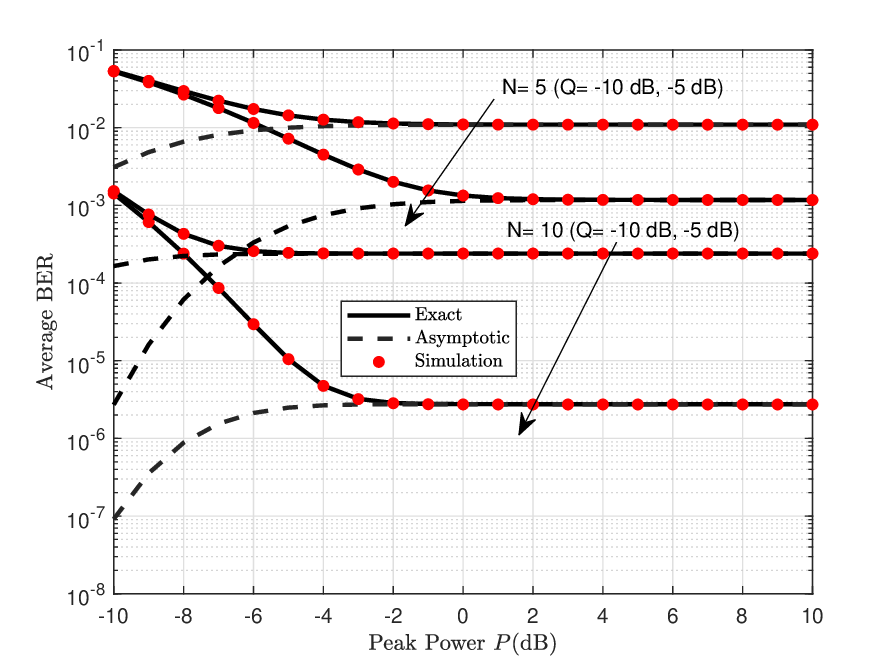}
	\caption{Average BER of the secondary network versus  the \ac{SS} peak power $P$  for various values of RIS elements $N$ at $\lambda= -5 \ \text{dB}$. }
	\label{fig_BER_Omega_N}
\end{figure}

In Fig. 5, we plot the ergodic capacity versus $N$ for various values of $Q$. We investigate the influence of increasing $N$ on the ergodic capacity and validate the accuracy of our exact and asymptotic expressions through Monte-Carlo simulations. The observations about the asymptotic curve align with our discussion in previous figures.

Fig. 6, illustrates the average \ac{BER} of \ac{BPSK} modulation format versus $P$ for different values of $N$. We first notice that the exact average \ac{BER} in (\ref{eq:L15}) matches perfectly with the Monte-Carlo simulations. It is evident from this figure that average \ac{BER} improves by either increasing $N$ or $Q$. Furthermore, we note that the asymptotic curves for the average \ac{BER}  tend to converge to the exact curves at large values of $P$.

\section{Conclusion} 
In this work, we proposed  a novel \ac{RIS}-assisted underlay spectrum sharing system, in which a secondary network assisted by an RIS shares the spectrum dedicated to a primary network. The performance of the \ac{RIS}-assisted secondary network over Rician fading channels is investigated while considering a practical transmit power adaption at the \ac{SS} and an interference power constraint at the \ac{PR}. Novel exact analytical expressions, in  terms of the $H$-function and the incomplete $H$-function, are derived to quantify the outage performance, the error performance, and the ergodic capacity.  Asymptotic expressions are also obtained for large values of the peak power of the \ac{SS}.  The derived results suggest that increasing the number of \ac{RIS} elements $N$ will substantially enhance the performance of the secondary network.  In addition, our analytical approach is validated through extensive Monte-Carlo simulations. 
\normalcolor

\color{black} 
\section*{Appendix A \\ Proof Of Theorem 1}
The \ac{CDF} of $\rho$ in (\ref{eq:L3}) can be derived as 
\begin{equation} \label{eq:A1}
\begin{split}
F_{\rho} (z) &= \Pr \{ \rho \leq z\} \\
&= \Pr \left\{  P R^{2} \leq z,  |h_{0}|^{2} \leq \frac{Q}{P} \right\} \\
&\quad + \Pr \left\{  \frac{Q R^{2}}{ |h_{0}|^{2}} \leq z,  |h_{0}|^{2} > \frac{Q}{P} \right\}. 
\end{split}
\end{equation}
 In view  of $|h_{0}|^{2}$ and $R^{2}$ being independent RVs, $F_{\rho}(z)$ can be evaluated as 
\begin{align} \label{eq:A2}
\begin{split}
F_{\rho} (z)&=  \Pr \left\{ R^{2} \leq \frac{z}{P} \right\}  \Pr \left\{  |h_{0}|^{2} \leq \frac{Q}{P} \right\}   \\
&\quad+  \int_{ \frac{Q}{P}}^{\infty} \Pr \left\{  R^{2} \leq \frac{z x}{Q} \right \}  f_{|{h}_{0}|^{2}} (x) \mathrm{d}x.  
\end{split}
\end{align}
Using the fact that $|{h}_{0}|^{2}$ is an exponential RV with mean $\lambda$,  then $F_{\rho} (z)$ can be expressed in terms of the CDF of $R^{2}$ as
\begin{align} \label{eq:A3}
\begin{split}
F_{\rho} (z)&=  F_{R^{2}} \left( \frac{z}{P}\right)  \left( 1- e^{-\frac{Q}{P \lambda}} \right)   \\
&\quad+  \int_{ \frac{Q}{P}}^{\infty} F_{R^{2}} \left( \frac{zx}{Q}\right) \frac{1}{\lambda} e^{-\frac{x}{\lambda}} \mathrm{d}x.  
\end{split}
\end{align}
Making use of (\ref{eq:L4}) in (\ref{eq:A3}) yields
\begin{align} \label{eq:A4}  
\begin{split}  
F_{\rho} (z)&= \frac{\gamma \left(v,  \sqrt{\frac{z}{b^2 P}} \, \right)}{\Gamma(v)}  \left( 1- e^{-\frac{Q}{P \lambda}} \right)   \\
&\quad+ \underbrace{ \int_{ \frac{Q}{P}}^{\infty} \frac{ \gamma \left(v,   \sqrt{\frac{zx}{b^2 Q}} \, \right)}{\Gamma(v)}  \frac{1}{\lambda} e^{-\frac{x}{\lambda}} \mathrm{d}x}_{I_{1}}.  
\end{split}
\end{align}

We focus next on evaluating the integral expression $I_{1}$ in (\ref{eq:A4}). The first step is to express the lower incomplete Gamma function in terms  of the Meijer $G$-function as \cite[Eq. (8.4.16.1)]{prudnikov1990integrals}   
\begin{equation} \label{eq:A5}
\begin{split}
\ \gamma \left(v,   \sqrt{\frac{zx}{b^2 Q}}\, \right)&=  G_{1,2}^{1,1}\left( \left.  \sqrt{\frac{zx}{b^2 Q}} \right| \begin{smallmatrix} 1 \\ ~\\ v, 0 \end{smallmatrix} \right) \\
&= \frac{1}{2\pi j} \int\limits_{\mathcal{L}}  \frac{ \Gamma(s)   \Gamma(v-s)  \left(\frac{zx}{b^2 Q}\right)^{\frac{s}{2}}}  {\Gamma(1+s)} \mathrm{d}s,
\end{split}
\end{equation}
where the second line in (\ref{eq:A5}) represents the Mellin-Barnes integral of the Meijer $G$-function \cite[Definition 1.5]{mathai2009h}. 

Plugging  (\ref{eq:A5}) into the integral expression of (\ref{eq:A4}) will result in  (\ref{eq:A7}). The inner integral in (\ref{eq:A7}) can be evaluated with the help of  \cite[Eq. (3.381.3)]{gradshteyn2014table} as 
	\begin{equation}  \label{eq:A6}
		\begin{split}
			I_{2}=\int_{\frac{Q}{P}}^{\infty} x^{\frac{s}{2}} ~ \frac{1}{\lambda} e^{-\frac{x}{\lambda}}    \mathrm{d}x = \lambda^{\frac{s}{2}} \,   \Gamma\left(1+\frac{s}{2}, \frac{Q}{P \lambda} \right).  
		\end{split}
	\end{equation}
Substituting (\ref{eq:A6}) into (\ref{eq:A7}) leads to (\ref{eq:A8}). Considering  (\ref{eq:A8}) and using the definition of the incomplete $H$-function \cite[Eq. (2.3)]{srivastava2018incomplete}, $F_{\rho} (z)$ in (\ref{eq:A4})  can be finally expressed in a compact form as in  (\ref{eq:L5}). 

To this end, $f_\rho(z)$ is obtained as shown in (\ref{eq:L6}) by differentiating $F_\rho(z)$ with respect to $z$  with the help of   \cite[Eq. (2.12)]{srivastava2018incomplete}.  This completes the proof of Theorem 1.

\begin{figure*}[t!]
\color{black} 
	\begin{equation}  \label{eq:A7}
		\begin{split}
I_{1} & =  \frac{1}{2\pi j  \Gamma(v)} \int\limits_{\mathcal{L}}  \frac{ \Gamma(s)   \Gamma(v-s)  \left(\frac{z}{b^2 Q}\right)^{\frac{s}{2}}}  {\Gamma(1+s)}  \underbrace{\left(\int_{\frac{Q}{P}}^{\infty} x^{\frac{s}{2}} ~ \frac{1}{\lambda} e^{-\frac{x}{\lambda}}    \mathrm{d}x \right) }_{I_2} \mathrm{d}s.\\
		\end{split}
	\end{equation}
\end{figure*}

\begin{figure*}[t!]
\color{black} 
	\begin{equation}  \label{eq:A8}
		\begin{split}
I_{1}= \frac{1}{2\pi j  \Gamma(v)} \int\limits_{\mathcal{L}}  \frac{ \Gamma\left(1+\frac{s}{2}, \frac{Q}{P \lambda} \right) \Gamma(s)   \Gamma(v-s)  \left(\frac{z \lambda }{b^2 Q}\right)^{\frac{s}{2}}}  {\Gamma(1+s)}   \mathrm{d}s. \\
		\end{split}
	\end{equation}
	\hrulefill
\end{figure*}

\begin{figure*}[t!]
\color{black} 
	\begin{equation}  \label{eq:A14}
		\begin{split}
 f^{(2)}_{\rho} (z)= \frac{z^{-1}}{2\pi j  \Gamma(v)} \int\limits_{\mathcal{L}}  \frac{ \Gamma\left(1+\frac{s}{2}, \frac{Q}{P \lambda} \right) \Gamma(s)   \Gamma(v-s) \Gamma(1+\frac{s}{2})  \left(\frac{z \lambda }{b^2 Q}\right)^{\frac{s}{2}}}  {\Gamma(1+s) \Gamma(\frac{s}{2})}   \mathrm{d}s. \\
		\end{split}
	\end{equation}
\end{figure*}

\begin{figure*}[t!]
\color{black} 
	\begin{equation}  \label{eq:A15}
		\begin{split}
\Ce^{(2)}= 
\frac{1}{2\pi j  \ln(2) \Gamma(v)} \int\limits_{\mathcal{L}}  \frac{ \Gamma\left(1+\frac{s}{2}, \frac{Q}{P \lambda} \right) \Gamma(s)   \Gamma(v-s) \Gamma(1+\frac{s}{2})  \left(\frac{\lambda }{b^2 Q}\right)^{\frac{s}{2}}}  {\Gamma(1+s) \Gamma(\frac{s}{2})} \underbrace{ \left( \int_{0}^{\infty} z^{\frac{s}{2}-1} \ln(1+z) \mathrm{d}z \right)}_{I_3} \mathrm{d}s. \\
		\end{split}
	\end{equation}
\end{figure*}

\begin{figure*}[t!]
\color{black} 
	\begin{equation}  \label{eq:A16}
		\begin{split}
\Ce^{(2)}= 
\frac{1}{2\pi j  \ln(2) \Gamma(v)} \int\limits_{\mathcal{L}}  \frac{ \Gamma\left(1+\frac{s}{2}, \frac{Q}{P \lambda} \right) \Gamma(s)   \Gamma(v-s) \Gamma(1-\frac{s}{2})   \Gamma(\frac{s}{2})  \left(\frac{\lambda }{b^2 Q}\right)^{\frac{s}{2}}}  {\Gamma(1+s) }  \mathrm{d}s. \\
		\end{split}
	\end{equation}
	\hrulefill
\end{figure*}

\section*{Appendix B \\ Proof Of Theorem 2}

The first step is to express $f_{\rho} (z)$ in  (\ref{eq:L6}) as 
\begin{equation} \label{eq:A9}  
f_{\rho} (z)= f^{(1)}_{\rho} (z)+f^{(2)}_{\rho} (z). 
\end{equation}
Here, we consider  $ f^{(1)}_{\rho} (z)= \frac{z^{\frac{v}{2}-1}  e^{- \sqrt{\frac{z}{b^2 P}}}     }{2(b^{2} P)^{\frac{v}{2}} \Gamma(v)}  \left( 1- e^{-\frac{Q}{P \lambda}} \right) $ and \\

\noindent $ f^{(2)}_{\rho} (z)= \frac{z^{-1}}{\Gamma(v)} \Gamma_{3,3}^{1,3}\left[ \left.  \sqrt{\frac{z\lambda}{b^2 Q}} \right| \begin{smallmatrix} (0,\frac{1}{2}, \frac{Q}{P \lambda} ), (0, \frac{1}{2}), (1, 1) \\ ~\\ (v, 1), (0, 1), (1, \frac{1}{2}) \end{smallmatrix} \right]. $

We now evaluate $\Ce^{(1)}$ as 
\begin{equation} \label{eq:A10}  
\Ce^{(1)}= \frac{1}{\ln(2)} \int_{0}^{\infty}  \ln \left( 1+ z \right)  f^{(1)}_{\rho} (z) \mathrm{d}z. 
\end{equation}
Using the transformation  of variables $u= \sqrt{\frac{z}{b^2 P}}$, the integral above can be expressed as  
\begin{equation} \label{eq:A11}  
\Ce^{(1)}= \frac{  \left( 1- e^{-\frac{Q}{P \lambda}} \right) }{\ln(2) \Gamma(v)} \int_{0}^{\infty}  \ln \left( 1+ b^{2} P u^{2} \right) u^{v-1} e^{-u}  \mathrm{d}u. 
\end{equation}
The term $ \ln \left( 1+ b^{2} P u^{2} \right)$ can written in terms of Meijer G-function as \cite[ Eq. (8.4.6.5) ]{{prudnikov1990integrals}} as
\begin{equation}\label{eq:A12}
\ln \left( 1+ b^{2} P u^{2} \right) = G_{2,2}^{1,2}\left( \left. b^{2} P u^{2}  \right| \begin{smallmatrix} 1, 1 \\ ~\\ 1,  0 \end{smallmatrix} \right).
\end{equation}
To this end, substituting  (\ref{eq:A12}) in  (\ref{eq:A11}) and applying \cite[Eq. (2.29) ]{mathai2009h}, $\Ce^{(1)}$ can be finally expressed as in (\ref{eq:L11}). 

In what follows, we evaluate $\Ce^{(2)}$ as 
\begin{equation} \label{eq:A13}  
\Ce^{(2)}= \frac{1}{\ln(2)} \int_{0}^{\infty}  \ln \left( 1+ z \right)  f^{(2)}_{\rho} (z) \mathrm{d}z. 
\end{equation}
We first utilize \cite[Eq. (2.3)]{srivastava2018incomplete} to expres  $f^{(2)}_{\rho} (z)$ in terms of the  Mellin-Barnes integral as in (\ref{eq:A14}). 

Plugging  (\ref{eq:A14}) into (\ref{eq:A13}) will result in  (\ref{eq:A15}). The inner integral in (\ref{eq:A15}) can be evaluated with the help of  \cite[Eq. (4.293.3)]{gradshteyn2014table} as 
	\begin{equation}  \label{eq:A17}
		\begin{split}
			I_{3}= \int_{0}^{\infty} z^{\frac{s}{2}-1} \ln(1+z) \mathrm{d}z= \frac{\pi}{\frac{s}{2} \sin \left( \frac{\pi s}{2}\right)}. 
		\end{split}
	\end{equation}
We now utilize \cite[Eq. (8.334.3) and Eq. (8.331.1)]{gradshteyn2014table} to express $I_{3}$ in terms of a product of gamma functions as
	\begin{equation}  \label{eq:A18}
		\begin{split}
			I_{3}= \frac{ \Gamma \left(1- \frac{s}{2} \right)  \Gamma \left( \frac{s}{2} \right) \Gamma \left( \frac{s}{2} \right) }{  \Gamma \left(1+ \frac{s}{2} \right) }. 
		\end{split}
	\end{equation}
Making use of (\ref{eq:A18}) in  (\ref{eq:A15}) leads to  (\ref{eq:A16}).  Considering  (\ref{eq:A16}) and using the definition of the incomplete $H$-function \cite[Eq. (2.3)]{srivastava2018incomplete}, $\Ce^{(2)}$ in (\ref{eq:A13})  can be finally expressed in a compact form as in  (\ref{eq:L12}).  This concludes Theorem 2.

\section*{Appendix C \\ Proof Of Theorem 3}
Consider the first and the second terms of $F_{\rho} (z)$ in  (\ref{eq:L5}) as  $F^{(1)}_{\rho} (z)$ and  $F^{(2)}_{\rho} (z)$, respectively. Then, $P^{(1)}_{e}$ can be expressed as
\begin{equation} \label{eq:A19}  
\begin{split}
		P^{(1)}_{e} &=\frac{\delta^{\zeta}}{2 \Gamma(\zeta)} \int_{0}^{\infty} e^{-\delta z} z^{\zeta-1} F^{(1)}_{\rho} (z) \mathrm{d}z  \\
&= \frac{\ \left( 1- e^{-\frac{Q}{P \lambda}} \right)   \delta^{\zeta} }{2 \Gamma(v) \Gamma(\zeta)} \int_{0}^{\infty} e^{-\delta z} z^{\zeta-1}{\gamma \left(v,  \sqrt{\frac{z}{b^2 P}} \, \right)}  \mathrm{d}z.   
\end{split}		
\end{equation}
The lower incomplete Gamma function in the second line of \ref{eq:A19}) can be expressed in terms  of the Meijer $G$-function as \cite[Eq. (8.4.16.1)]{prudnikov1990integrals}   
\begin{equation} \label{eq:A20}
\begin{split}
\ \gamma \left(v,   \sqrt{\frac{z}{b^2 P}}\, \right)&=  G_{1,2}^{1,1}\left( \left.  \sqrt{\frac{z}{b^2 P}} \right| \begin{smallmatrix} 1 \\ ~\\ v, 0 \end{smallmatrix} \right). 
\end{split}
\end{equation}
Making use of  (\ref{eq:A20}) in  (\ref{eq:A19}) and utilizing the Laplace transform property of  the Meijer $G$-function  \cite[Eq. (2.29) ]{mathai2009h}, $P^{(1)}_{e}$ is finally as given in  (\ref{eq:L16}). 

Similarly,  $P^{(2)}_{e}$ can be obtained from  $F^{(2)}_{\rho} (z)$ as    
\begin{equation} \label{eq:A21}
\begin{split}  
		P^{(2)}_{e}& =\frac{\delta^{\zeta}}{2 \Gamma(\zeta)} \int_{0}^{\infty} e^{-\delta z} z^{\zeta-1} F^{(2)}_{\rho} (z) \mathrm{d}z\\
&=\frac{  \delta^{\zeta} }{2 \Gamma(v) \Gamma(\zeta)} \int_{0}^{\infty} e^{-\delta z} z^{\zeta-1}  \\
&\quad  \times \Gamma_{2,2}^{1,2}\left[ \left.  \sqrt{\frac{z\lambda}{b^2 Q}} \right| \begin{smallmatrix} (0, \frac{1}{2}, \frac{Q}{P \lambda} ), (1, 1) \\ ~\\ (v, 1), (0, 1) \end{smallmatrix} \right] \mathrm{d}z.
		\end{split}   
\end{equation}
Considering (\ref{eq:A21}) and utilizing the Laplace transform property of the incomplete $H$-function \cite[Eq. (3.6)]{srivastava2018incomplete}, $P^{(2)}_{e}$ is finally as given in  (\ref{eq:L17}).  This completes the proof of Theorem 3.

\normalcolor

\bibliographystyle{IEEEtran}
\bibliography{main}

\begin{IEEEbiography}[{\includegraphics[width=1in,height=1.2in,clip]{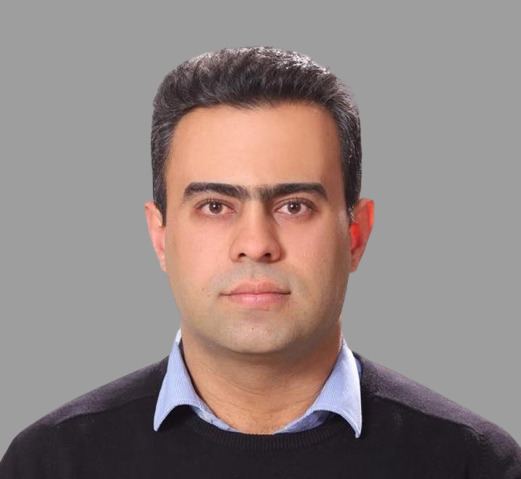}}]{Yazan H. Al-Badarneh} was born in Irbid, Jordan, in 1987. He obtained his Ph.D. in Electrical Engineering from Texas A\&M University, College Station, TX, USA, in 2018. Following his graduation, he joined the Department of Electrical Engineering at The University of Jordan, Amman, Jordan, as an Assistant Professor. In September 2023, in recognition of his academic contributions, he was promoted to the rank of Associate Professor within the same department. Since September 2023, he has been serving as a Deputy Dean for Quality Affairs, Accreditation, and Ranking at the School of Engineering, The University of Jordan. His scholarly pursuits predominantly focus on the application of information and communication theories to contemporary wireless communication systems, with particular emphasis on modeling, design, and performance analysis.
\end{IEEEbiography}
\vskip 0pt plus -1fil

\begin{IEEEbiography}[{\includegraphics[width=1in,height=1.2in,clip]{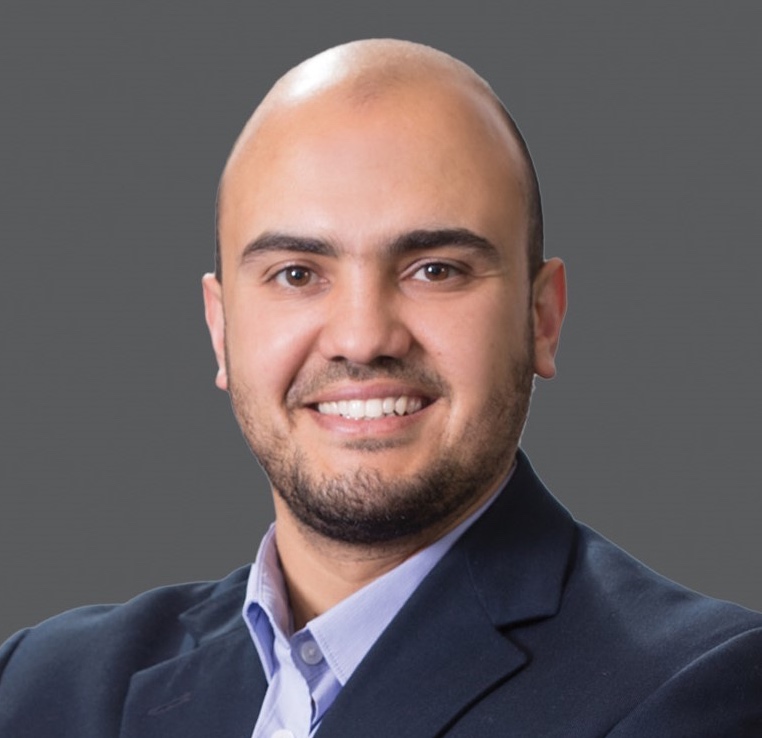}}]{Mustafa Alshawaqfeh} received his Ph.D. degree in Electrical Engineering from Texas A\&M University, College Station, Texas, USA, in 2017, M.Ss., degree in wireless communication from the Jordan University of Science and Technology, Irbid, Jordan, in 2010, and B.Ss. degree in communication engineering from Yarmouk University, Irbid, Jordan, in 2007.  He joined the German Jordanian University in 2017 as an Assistant Professor.  In September 2021, He was promoted to an Associate Professor in the EE department. His research interests span the areas of wireless communications, signal processing, machine learning, and bioinformatics. In the area of wireless communication, he worked on (i) applying signal processing, mainly sparse recovery and tree search algorithms, to develop efficient low complexity receivers especially for spatial modulation schemes and non-orthogonal multiple access (NOMA) systems, and (ii) analyzing the performance of wireless MIMO systems over fading channels.  In the area of bioinformatics, he is working on applying signal processing and machine learning techniques to analyze biological datasets with the focus on developing algorithms for metagenomic biomarker detection. 
\end{IEEEbiography}
\vskip 0pt plus -1fil 


\begin{IEEEbiography}[{\includegraphics[width=1in,height=1.25in,clip,keepaspectratio]{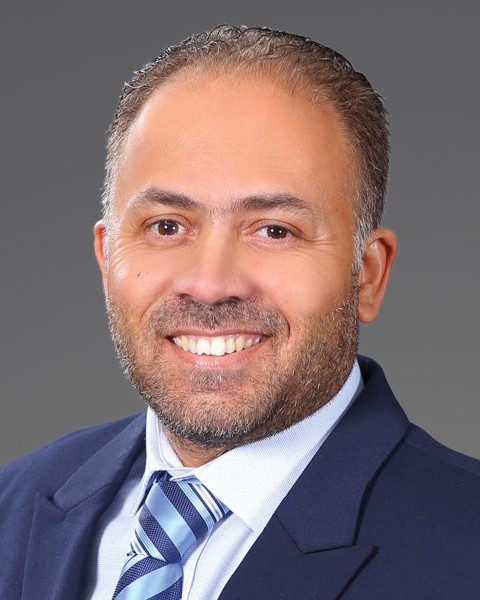}}]{Osamah S. Badarneh} received his PhD in Electrical Engineering from the University of Quebec-\'{E}cole de Technologie Sup\'{e}rieure (ETS), Canada, in 2009. He joined the German Jordanian University (GJU) in 2018 and he is currently a Full Professor with the Department of Electrical Engineering. He served as an Adjunct Professor with the Department of Electrical Engineering, University of Quebec-ETS, 2013-2018. From 2012 to 2018, he was an Associate Professor with the Department of Electrical Engineering, University of Tabuk. Also, he worked as an Assistant Professor with the Department of Telecommunication Engineering, Yarmouk University, from 2010 to 2012. His research interests focus on wireless communications and networking. He is the author of more than 100 publications in scientific journals and international conferences.
\end{IEEEbiography}
\vskip 0pt plus -1fil 

\begin{IEEEbiography}[{\includegraphics[width=1in,height=1.25in,clip,keepaspectratio]{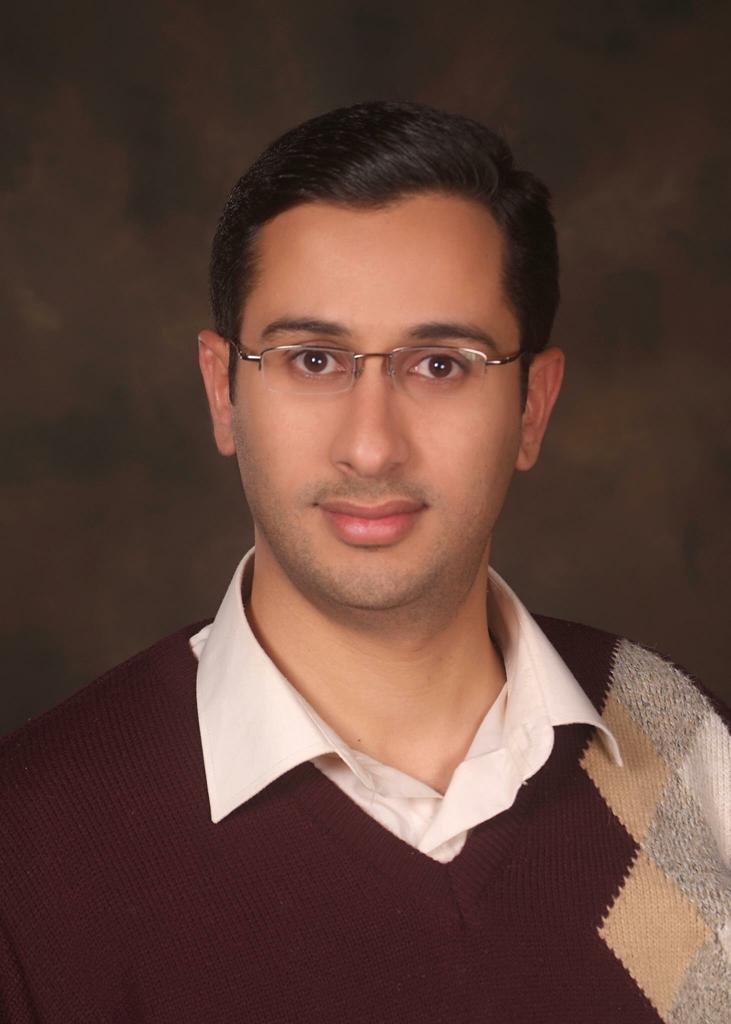}}] {Yazid M. Khattabi} received the bachelor's degree in electrical engineering with a specialization in electronics and communications and the master's degree in electrical engineering with a specialization in wireless communications both from Jordan University of Science and Technology (JUST), Irbid, Jordan, in 2008 and 2010, respectively, and the Ph.D. degree in electrical engineering with a specialization in wireless communications from the Center for Wireless Communications (CWC), University of Mississippi, Oxford, MS, USA, in 2016. From March 2011 to December 2012, he served as a telecommunications and electronics design engineer at King Abdullah II Design and Development Bureau (KADDB), Amman, Jordan, with responsibilities included and not limited to telecommunication systems design for unmanned vehicles; electronics and RF printed circuits software design, implementation, and maintenance; RF planning \& optimization; signal propagation modeling; etc. From 2013 to 2016, he served as a research assistant with the CWC, University of Mississippi, where he received several research awards. Since August, 2016, he has been with the University of Jordan, Amman, Jordan, where he is currently an Associate Professor of electrical engineering. His current research interests include several areas in wireless communications with emphasis on performance evaluation and optimization of backscatter communications, reconfigurable intelligent surfaces (RIS) communications, relaying and cooperative communications, cognitive radio based communications, MIMO and spatial modulation (SM) communications, high-speed-railway communications, wireless powered communications, internet of things (IoT) communications, lightweight encryption algorithms, vehicular and unmanned-aerial-vehicle (UAV) communications, mm-wave and Tera-Hz communications. He is also serving as a reviewer for several refereed international journals and conferences.s
\end{IEEEbiography}

\end{document}